\documentclass[twocolumn,showpacs,preprintnumbers,amsmath,amssymb,prb,aps]{revtex4}

\usepackage{epsfig} %Include figure files
\usepackage{graphicx}% Include figure files
\usepackage{dcolumn}% Align table columns on decimal point
\usepackage{bm}% bold math
\usepackage{epstopdf}
\usepackage{amsmath}
\begin{document}

\title{Structural response to the magnetic pre ordering in LiFeSi$_2$O$_6$}
\author{R. K. Maurya$^1$}
\author{Priyamedha Sharma$^1$}
\author{R. Rawat$^2$}
\author{R. S. Singh$^3$}
\author{R. Bindu$^1$}
\altaffiliation{Corresponding author: bindu@iitmandi.ac.in}
\affiliation{$^1$School of Basic Sciences, Indian Institute of Technology Mandi, Kamand, Himachal Pradesh- 175005, India}
\affiliation{$^2$ Inter University Consortium for Departmnet of Atomic Energy Facilities, University Campus, Khandwa Road,Indore-452017, India}
\affiliation{$^3$ Department of Physics, IISER Bhopal, Bhopal-462066, India}

\date{\today}
\begin{abstract}
We investigate the temperature evolution of the structural parameters of potential ferrotoroidic LiFeSi$_2$O$_6$ compound across structural and magnetic phase transitions. The structural transition (T$_S$)is around 220K and the paramagnetic to antiferromagnetic transition (T$_N$) is around 18K. The lattice parameters exhibit unusual temperature dependence and based on its behaviour, the experimental results can be divided into 3 regions.In region I (300K to 240K), the cell parameters are mainly governed by mere thermal effect. As the compound enters region II (below 240K to 50K), the lattice parameters show non linear behaviour. In this region, the exchange pathways that lead to the magnetic interactions within and between the Fe-Fe chains do not show significant response.The region III (below 50K) is dominated by the magnetic contribution where we observe setting up of intra and inter-chain magnetic interaction. This behaviour is unlike other low dimensional compounds like Ca$_3$Co$_2$O$_6$, Sr$_3$NiRhO$_6 $, MnTiO$_3$ etc. thereby suggesting the magnetism in LiFeSi$_2$O$_6$ is of three dimensional nature. The present results will be helpful in understanding the evolution of the spin rings that give rise to net toroidal moment and hence its multiferroic behaviour.    
\end{abstract}


\maketitle
\section{Introduction}
Magnetism, ferroelectricity and superconductivity are some of the quantum phenomena that is manifested at the macroscopic scale and we experience in our real life. In strongly correlated electron systems, these properties arise because of the intricate coupling among the charge, spin, orbital and lattice degrees of freedom. The materials in which there is coexistence of two or more ferroic orders are known as the multiferroics \cite{Schmid}. The interesting aspect of these kinds of materials is that one can obtain not only its parent property but also the properties that arises due to its cross coupling between the different ferroic orders that leads to the behaviour of multi-functionality \cite{Wang}. Such behaviour is in high demand with the increase in device maniaturization \cite{Yang}. Recently, coexistence of ferroelectricity and antiferromagnetism (AFM) has drawn great attention both from fundamental physics point of view and also its application in high density memory devices and magnetic field sensor \cite{Schmid}. These kinds of systems exhibit close coupling between spin and lattice degrees of freedom. LiFeSi$_2$O$_6$ is one such compound that displays ferrotoroidicity, a new class of primary ferroic order in which the toroidal moment align parallel spontaneously and antiferromagnetism with Neel temperature around 18K \cite{Jodlauk, Baum, Toledano}.Ferrotoroidicity involves cross product of space part and magnetic part. In this material, the ferrotoroidicity is driven by the magnetic field \cite{Toledano}.\\

\begin{figure}
	\vspace{-1ex}
	\includegraphics [scale=0.4, angle=0]{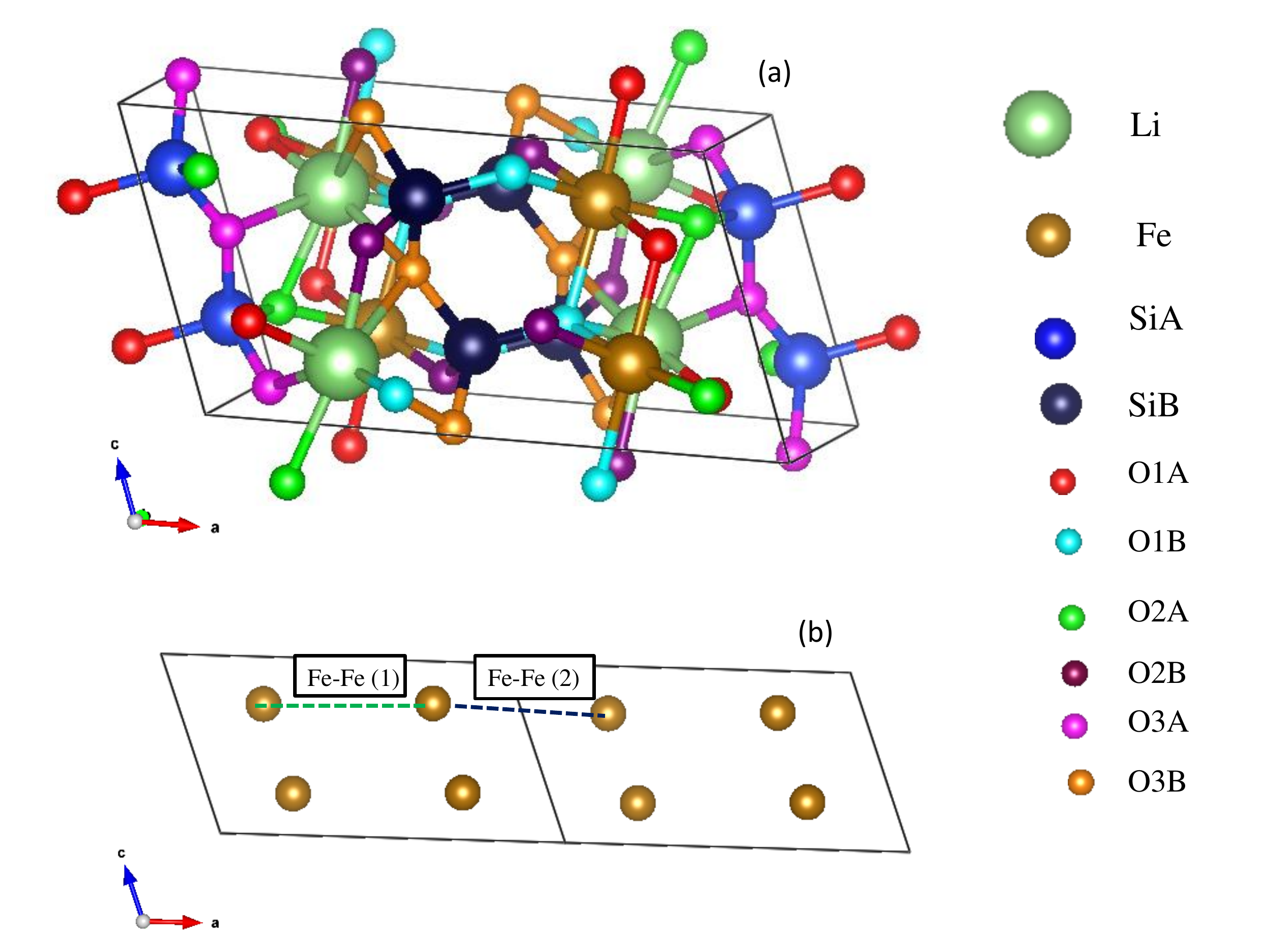}
	\vspace{-5ex}
	\caption{(a) crystal structure of LiFeSi$_2$O$_6$; (b) The intra chain Fe-Fe bonds (Fe-Fe(intra)) and inter chain Fe-Fe bonds namely Fe-Fe(1) and Fe-Fe(2) are depicted pictorially.}
	\vspace{-2ex}
\end{figure}

LiFeSi$_2$O$_6$, the compound under study belongs to pyroxene family of AMSi$_2$O$_6$ type (A= mono or di valent metal and M=di or tri valent metal). At room temperature, this material stabilizes in the monoclinic structure with the space group C2/c and stabilizes in P2$_1$/c space group of monoclinic around 230K \cite{Redhammer}. In this compound, the FeO$_6$ octahedra are edge shared along the c-axis forming zig-zag chains and each of these chains are connected to each other through SiO$_4$ octahedra. Magnetically, these FeO$_6$ chains are ferromagnetic and between the chains, it is antiferromagnetic interaction.Because of the formation of FeO$_6$ chains, it is expected to show quasi one dimensional behaviour. The spontaneous toroidal moments arise due to the formation of spin rings. However, the formation of toroidal moment demands that the exchange interaction between the Fe ions that forms a spin ring should have the same strength \cite{Changhoon}. Apart from these, there are reports that mention about magnetic pre-ordering but the information about clear onset temperature of the magnetic interaction is lacking. To understand about the structural connectivity with the magnetsim and dimensionality and the magnetic pre-ordering temperature, we have carried out temperature dependent x-ray diffarction experiments on LiFeSi$_2$O$_6$. Our results show the clear connection between the structural parameters and the pre-prdering of magnetic interactions. The magnetism existing in this compound appears to be of three dimensional character.

\begin{figure}
	\vspace{-1ex}
	\includegraphics [scale=0.4, angle=0]{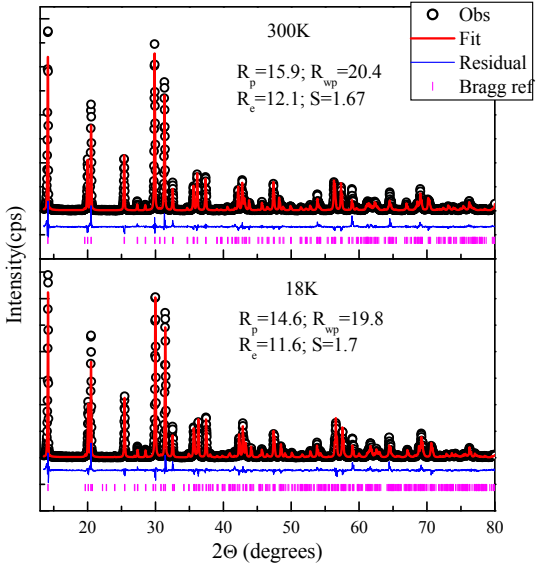}
	\vspace{-20ex}
	\caption{(a) Rietveld refinement of the xrd patterns of LiFeSi$_2$O$_6$ collected at (a) 300 K and (b) 18 K.}
	\vspace{-2ex}
\end{figure}

\section{Results and Discussions}
In Fig.1(a), we show the crystal structure of LiFeSi$_2$O$_6$ compound. The Rietveld refinement of the x-ray diffraction patterns of this compound collected at 300K and 18K are displayed in Fig.2. Our results show that down to low temperatures, the compound remains in the monoclinic crystal system but there occurs phase transition from C2/c to P2$_1$/c space group around 230K (T$_S$) \cite{Roth}.\\

The temperature dependent dc magnetic susceptobility at an external magnetic field of 0.5T is shown in Fig.3(a). The susceptibility curve reaches a maximum around 18K. This marks the Neel temperature (T$_N$). The field dependent magnetization plot shows S-shape curve suggesting spin flop transition, inset of Fig.3(a).

\begin{figure}
	\vspace{-1ex}
	\includegraphics [scale=0.4, angle=0]{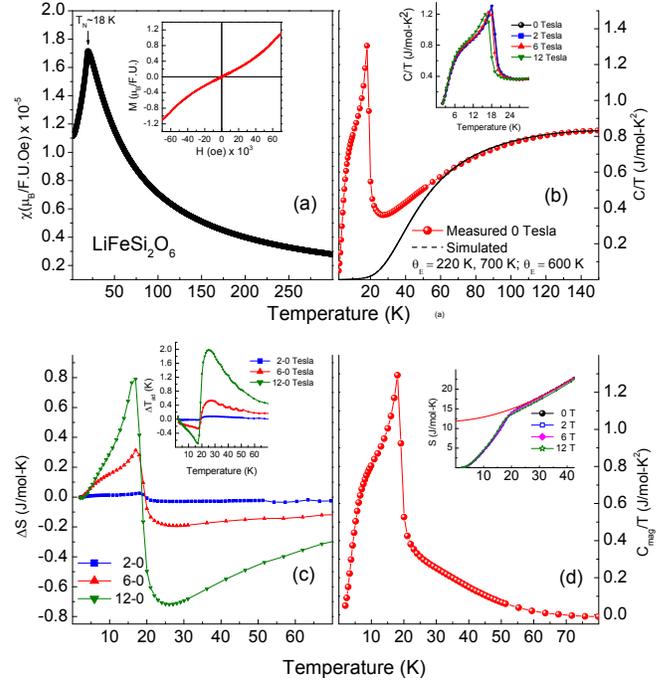}
	\vspace{-10ex}
	\caption{(a) The temperature evolution of the dc magnetic susceptibility at an applied magnetic field of 0.5 T; the inset shows magnetisation as a function of applied magnetic field collected at 4 K (b) The temperature dependence of heat capacity C$_p$ plotted in the form of C$_p$/T vs T. Black curve is simulated lattice contribution. Inset highlights the region close to transition temperature; (c)Isothermal entropy change ($\Delta$S$_t$$_{h}$) as a function of temperature; the inset shows adiabatic temperature change as function of temperature; (d) magnetic contribution to heat capacity in the form C$_m$$_a$$_g$/T vs T. Inset shows the temperature dependence of zero field total entropy (S$_t$$_o$$_t$$_a$$_l$) calculated by integrating the measured C$_p$/T vs. temperature curve.}
	\vspace{-2ex}
\end{figure}

The temperature dependent heat capacity C$_P$ plotted in the form of C$_P$/T vs T is shown in Fig.3(b). The inset of Fig. 3(b) shows the heat capacity collected at different applied magnetic fields (2,6 and 12T). A peak is observed around 18K and shifts to low temperature with the application of magnetic field of 6 and 12 T as highlighted in the inset of Fig. 3(b). This behaviour suggests the nature of the ordering is antiferromagnetic which becomes more evident in the magnetocaloric curve shown in Fig.3(c). These magnetocaloric curves i.e. (a) isothermal entropy chnage ($\Delta$S$_th$) as well as (b) adiabatic temperature change ($\Delta$T$_ad$) are calculated from the measured heat capacity data shown in Fig. 3(c). For a paramagnetic system $\Delta$S$_th$ is expected to be negative as application of magnetic field results in reduction of spin disorder. In the case of antiferromagnetic system, the application of magnetic field creates more disorder as it is against the ordering effect of exchange intercation. Therefore, $\Delta$S$_th$ is expected to be positive in the antiferromagnetic state and across the antiferro to paramagnetic transition, a sign reversal will be observed \cite{Rawat}. Fig. 3(c) shows sign reversal around the magnetic ordering temperature giving a clear signature of antiferromagnetic like order. As expected the magnitude of both $\Delta$S$_th$ and $\Delta$T$_ad$ are small.\\
To separate the magnetic contribution from the measured heat capacity, lattice contribution is simulated with two Einstein temperatures (220K and 700K) and one Debye temperature (600K) as has been done in earlier studies \cite{Baker}. The simulated cirve is shown as black curve in Fig. 3(b). The magnetic contribution taken as the difference of simulated C/T and measured C/T (0 Tesla) curves is shown in Fig. 3(d). It shows non-zero contribution well above transition temperature.

\begin{figure}
	\vspace{-1ex}
	\includegraphics [scale=0.4, angle=0]{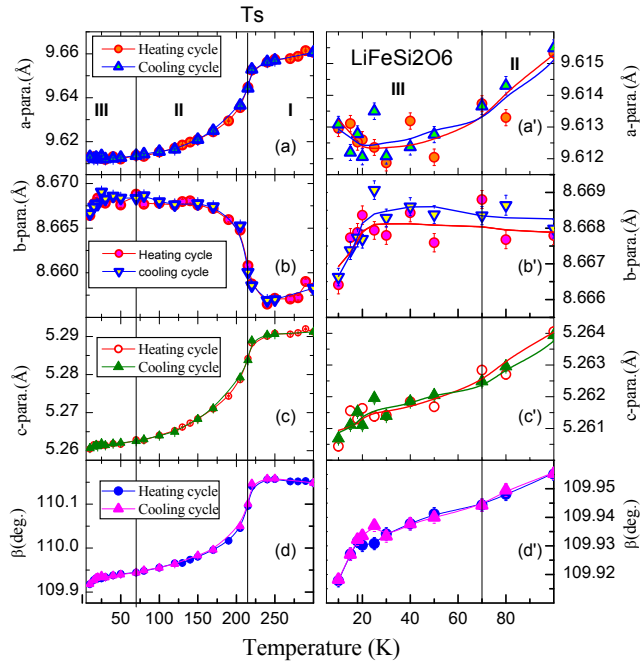}
	\vspace{-17ex}
	\caption{The first column shows the temperature behaviour of the lattice parameters where $\beta$ is the monoclinic angle; the second column shows temperature dependent lattice parameters in the temperature range 5 to 100 K. }
	\vspace{-2ex}
\end{figure}

The entropy associated with the transition is found to be about 14J/mol K around 20K(near T$_N$) which reaches to 20J/mol K at ~ 60K (where C$_mag$ reaches zero). For S=5/2 system it is expected to 14.89J/mol K. This is close to what has been observed near T$_N$. However, magnetic contribution is present up to ~ 60K, giving rise to much larger entropy change. This could be due to incorrect lattice contribution estimate as modeling with multiple Einstein and Debye function is not unique. To circumvent this ambiguity, we considered another approach. We calculated the total entropy from the measured zero field heat capacity data, which is shown as an inset to the Fig. 3(d). The entropy curve above the transition temperature is fitted with a third order polynomial (in the temperature range 30-90K) and extrapolated down to zero K. This intercept is found to be 12J/mol K which can be considered as the entropy of transition. This is about 80\% of that expected for the spin 5/2 system and is consistent with earlier reports \cite{Baker}.\\

The above results show the onset of magnetic contribution well above T$_N$. To understand this aspect, temperature dependent x-ray diffraction experiments were carried out.The lattice parameters obtained during the heating and cooling cycles from the Rietveld refinement are shown in Fig. 4. From the figure, we observe that with decrease in temperature until 240K, all the lattice parameters decrease linearly. As the compound enters T<T$_S $, the a and c parameters decreases while the b-parameter show an increase. Below T$_S$, the lattice parameters exhibit non-linear behaviour. The changes in the lattice parameters in the temperature region, below 240K to 50K are significant. In the temperature range, below 50 K to the lowest recorded temperature the a parameter remains almost the same and increases below T$_N$; the b-parameter remains almost the same and decreases below T$_N$ and the c-parameter is found to decrease, Figs. 4(a$'$-c$'$). Based on the behaviour of the cell parameters, we have divided the structural parameters in three regions.

Regions I, II and III are the temperature ranges 300K to 240K; below 240K to 50K and below 50K, respectively. In region I, the lattice parameters a, b and c decrease by 0.046\%, 0.02\% and 0.02\%, respectively. In region II, the changes in the lattice parameters are significant. The a and c parameters decrease by 0.44\% and 0.5\%, respectively and b-parameter increases by 0.11\%. In region III, the a nd b parameters remain almost the same unti around T$_N$ and later on these parameters show opposite behaviours. The c-parameter decreases down to the lowest collected temperature, Figs. 4(a$'$-c$'$). These results clearly show that there occurs sigificant change in the lattice parameters around T$_S$ and T$_N$. To understand these results, we look into the other structural parameters which also play significant role in deciding the magnetism within the Fe chain and between the Fe chains. In this paper, the thermal behaviour of the sructutal parameters are explained as a function of decrease in temperature.

In Fig, 5, we show the temperature variation of Fe-O bond lengths. For T>T$_S$, there are 3 types of oxygen ions, O1 (apical), O1 (basal) and O2 (basal). The apical and basal are referred to those oxygen ions that lie along the c-axis and in the ab-plane, respectively. For T<T$_S$, there are 6 types of oxygen ions. The O1 (apical) is split into O1A (apical) and O1B (apical); O1(basal) is split into O1A(basal) and O1B(basal) and O2(basal) is split into O2A and O2B. In region I, the temperature variation in the Fe-O bond distances remains almost the same. As the sample enters region II, the Fe-O1A apical bond becomes longer than Fe-O1B apical bond and later on no significant variation in Fe-O1A and Fe-O1B bonds occurs.
In the case of Fe-O1A (basal) and Fe-O1B(basal), there occurs slight increment and decrement, respectively. In the case of Fe-O2A and Fe-O2B (basal) bonds, there occurs significant increment and decrement, respectively. The Fe-O2A and B bonds decrease and increase, respectively. In the region III, Figs. 5(a$'$-c$'$), the Fe-O1A (apical), Fe-O1A(basal) and Fe-O2B are found to decrease and the Fe-O1B (apical), Fe-O1B (basal) and Fe-O2A show opposite behaviour. The Fe-O1A,B (apical) bond lengths are found to cross around T$_N$, Fig 5(a$'$). Similar behaviour is also observed in the case of Fe-O2A,B bond lengths, Fig. 5(c$'$). In the case of Fe-O1A,B (basal), they become almost identical around 10K, Fig. 5(b$'$).

\begin{figure}
	\vspace{-1ex}
	\includegraphics [scale=0.4, angle=0]{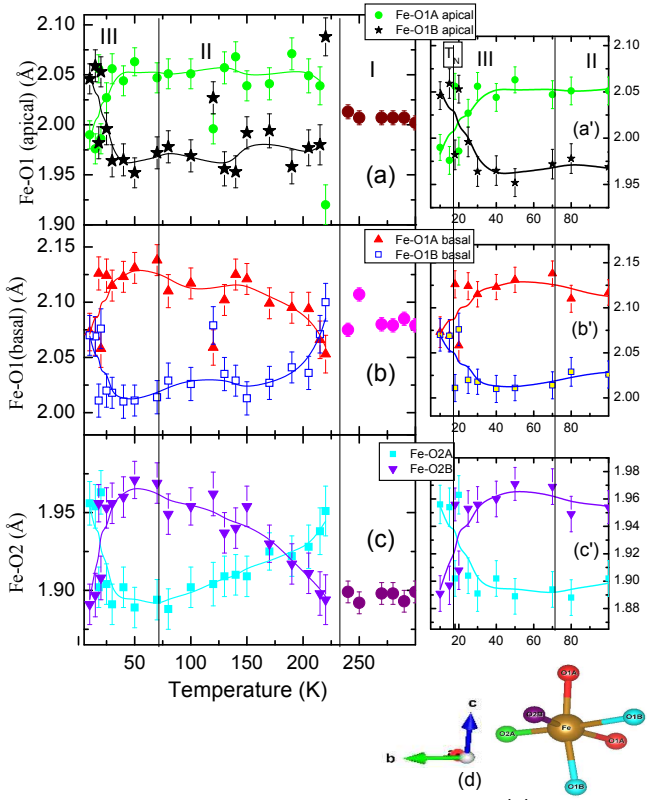}
	\vspace{-7ex}
	\caption{The first column shows the temperature behaviour of (a)Fe-O1 (apical) (b) Fe-O1 (basal) and (c) Fe-O2; the second column shows temperature dependent (a$'$)Fe-O1 (apical) (b$'$) Fe-O1 (basal) and (c$'$) Fe-O2 in region III; (d) the pictorial representation of FeO$_{6}$ octahedra in the P2$_{1}$/c phase.}
	\vspace{-2ex}
\end{figure}

In Fig. 6, we show the temperature variation of Si-O bonds. For T>T$_S$, there is only one kind of Si ion. For T<T$_S$, there are two kinds of Si ions labeled as SiA and SiB. In region I, there is no significant variation in the Si-O1 and Si-O2 bonds while the Si-O3(u) (u means 'up' along c-axis) bond shows marginal increment and then decrement and Si-O3(d)(d means 'down' in c-axis) shows opposite behaviour. As the sample enters region II, the Si-O1 bond is split into SiA-O1A and SiB-O1B. The SiB-O1B bonds are longer than the Si-O1A. The variation in these bond distances remain almost the same in this region. The Si-O2 bonds are split into SiA-O2A and SiB-O2B. The value of SiB-O2B is more than SiA-O2A. In this region, SiB-O2B bonds decrease by 1.2\% and SiA-O2A bonds increase by 0.7\%, respectively. The Si-O3(u) is split into SiA-O3A(u) and SiB-O3B (u). In this region, the values of both these bond distances are almost the same. The Si-O3(d) is split into SiA-O3A (d) and SiB-O3B (d). These bond distances are almost the same in this region. In region III, the Si-O1A increases by ~ 11\% and SiB-O1B decreases by ~ 10.8\%. The SiB-O2B decreases by ~ 3\% and SiA-O2A increases by ~ 0.9\%. The SiA-O3A(u) decreases by ~ 2.7\% and SiB-O3B(u) increases by ~2\%. The SiB-O3B(d) bond distances increases by ~ 0.8\% and SiA-O3A(d) bond distances remains unaltered.

\begin{figure}
	\vspace{-1ex}
	\includegraphics [scale=0.4, angle=0]{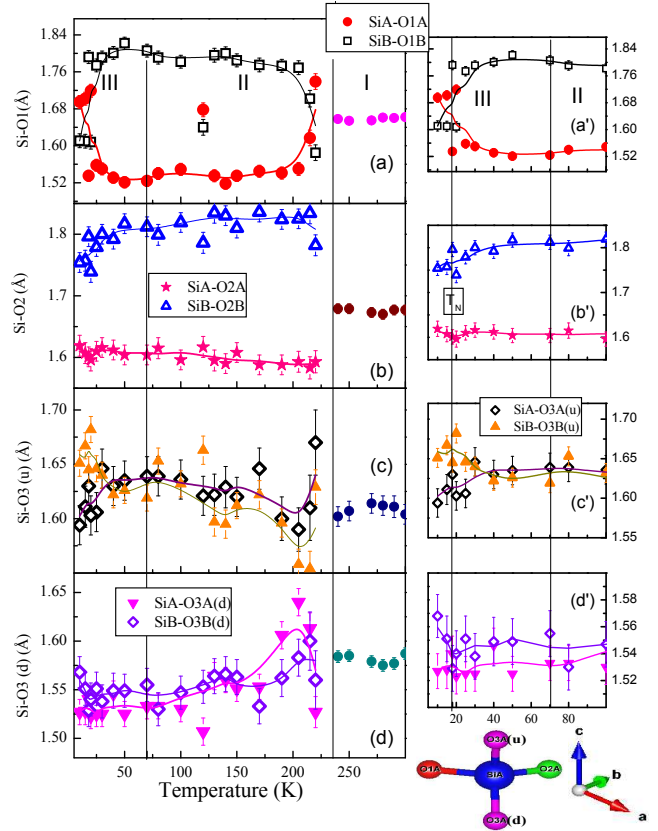}
	\vspace{-7ex}
	\caption{The first column shows the temperature behaviour of (a)Si-O1 (apical) (b) Si-O2 and (c) Si-O3(u) and (d) Si-O3(d) bonds; the second column shows temperature dependent (a$'$)Si-O1 (b$'$) Si-O2 (c$'$) Si-O3(u) and (d$'$) Si-O3(d) bonds in region III and (e) the pictorial representation of SiO$_{4}$ tetrahedra in the P2$_{1}$/c phase.}
	\vspace{-2ex}
\end{figure}

In Fig. 7(a-b), we show temperature behaviour of the Fe-Fe bonds. The Fe-Fe (intra) constitutes the bond distance between the edge shared FeO$_6$ octahedra (within the Fe chain) and Fe-Fe(1$' $) cinstitutes the bond distance between the Fe chains which is separated by SiO$_4$ tetrahedra. In region I, the Fe-Fe(intra) bond decreases by ~ 0.12\%. The Fe-Fe(1$'$) decreases by ~ 0.05\%.
When the sample enters region II, the Fe-Fe(1$'$) splits into Fe-Fe(1) and Fe-Fe(2). These bonds are separated by SiB and SiA tetrahedra, respectively. The Fe-Fe(1) bond is longer than Fe-Fe(2). In this region, all the Fe-Fe bonds within and between the chains remain almost the same. In region III, the Fe-Fe(1) bonds and Fe-Fe(2) bonds increase and decrease by ~ 0.82\% and ~ 0.77\%, respectively. The Fe-Fe(intra) bonds show slight decrement below 25K by ~ 0.16\%.

\begin{figure}
	\vspace{-1ex}
	\includegraphics [scale=0.4, angle=0]{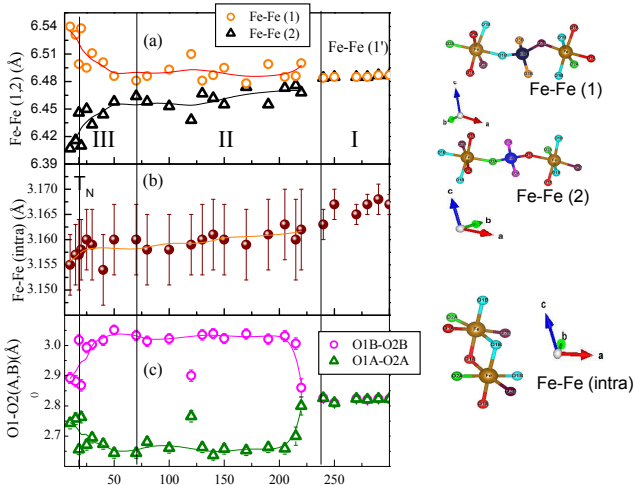}
	\vspace{-27ex}
	\caption{The first column shows the temperature behaviour of (a) Fe-Fe (1,2) (b) Fe-Fe (intra) (c) O1-O2 (A,B) bonds and the second column shows the pictorial representation of these bonds.}
	\vspace{-2ex}
\end{figure}

The above results show that the structural parameters exhibit significant changes in the region III which is below 50K. This behaviour is in line with the heat capacity results where the entropy change with regard to the magnetic interaction was found to be sufficient only in the temperature range 60K to 5K. The changes are observed in the Fe-O and Fe-Fe bond distances (between the chains).Toledano et al.\cite{Toledano} have shown the spin exchange pathway depicting the intra-chain ferromagnetic(Fe-Fe(intra)) and inter-chain intiferromagnetic (Fe-Fe (1)) interaction, Fig. 1(b) that form a spin ring. With the application of magnetic field, the spin ring is expected to give finite toroidal moment. In the region II, we observe discernible changes in the basal Fe-O bond distances. These changes are in such a way to preserve the volume of the FeO$_6$ octahedra.

The unaltered Fe-Fe intra-chain bond distances suggest insignificant role played by the direct exchange d-d orbital exchange of the Fe ions. This result is also in line with the prediction of Streltsov and Khomskii \cite{Streltsov}. They have shown that as one goes from light to heavy transition metal ion i.e. from Ti to Fe ions, the hopping integral (t$_dd$) between the direct d-d orbitals decrease provided the distance (D) between the transition metal ions are identical. The hopping integral t$_dd$ is given as rd3/d5, where r is the radius of the d states \cite{Harrison}. The inter-chain Fe-Fe(1) bonds are connected to the SiB-O4 tetrahedra as shown in Fig.7. Each of the Fe is connected to the SiB through O1B(basal) and O2B(basal). The O1B-O2B bond distances remain almost the same in region II due to rearragement of the Fe-O1B and Fe-O2B bonds.

In region III, we observe decrement in the O1B-O2B bonds suggesting that super-super exchange interaction between the chains dominates over the thermal effect. In this region, we also observe slight decrement in the Fe-Fe(intra) and the av. Fe-O bond remains almost the same. In the case of edge shared FeO$_6$ octahedra, antiferromagnetic interaction is predicted to set in, if the Fe-Fe bond angle is more than 97 degrees. It has has been observed in literature \cite{Redhammer,Streltsov}, the Fe-O1A-Fe(apical) and Fe-O1B-Fe(apical) bond angles are around 97.7 and 99.1 degrees at 1.4K. Hence the exchange pathway for the setting up of the ferromagnetic interaction within the chain is along Fe-O1A-Fe bond angle. We also observe significant decrement in the Fe-O1A(apical) bond distances in this region where ferromagnetic interaction occurs through O 2p orbitals. Apart from these behaviours, significant changes in the SiB-O1B and SiB-O2B bond distances are also observed. This suggests the possible involvement of Si 3s/3p orbitals in the setting up of super-super exchange interaction between the chains. Our results suggest that the onset of both the antiferromagnetic and ferromagnetic interactions occur around the same temperature and thus the magnetism is of three dimensional nature. The results obtained in the present study are different from the compounds that show low dimensional magnetism. In the case of quasi one dimensional Ca$_3$Co$_2$O$_6$\cite{Bindu} and Sr$_3$NiRhO$_6$ compounds\cite{Navneet}, from our EXAFS studies, through the behaviour of the local structural parameters, reports have shown clear signature of the onset of the ferromagnetic interactions (within the chain) at higher temperatures as compared to the onset of the antiferromagnetic interaction (between the chain). In the case of Ca$_3$Co$_2$O$_6$ compound, Mossbauer studies have also shown similar behaviour\cite{Paulose}. Similarly, in the case of quasi two dimensonal MnTiO$_3$ compound, our structural studies have shown clear signature of the onset of the magnetism within the ab-plane stabilizing at higher temperature than the onset of the magnetism between the planes\cite{Maurya}.
\section{Summary}
Temperature dependent x-ray diffraction experiments were carried out on LiFeSi$_2$O$_6$ compound. We observe interesting evolution of the structural parameters across the structural and magnetic phase transitions. The behaviour of the structural parameters reveal signature of magnetoc pre-ordering at temperatures well above T$_N$. Our results show that both antiferromagnetic interaction between the chain and ferromagnetic interaction within the chain order below 50K. Hence the magnetism existing in this compound is of three dimensional character. We believe that our results will be helpful in understanding the evolution of the spin rings that give rise to net toroidal moment if x-ray diffraction experiments are carried out in the presence of magnetic field.

\end{document}